\documentclass[prb,twocolumn,amssymb,superscriptaddress]{revtex4}
\usepackage{amsmath} 
\usepackage{amssymb} 
\usepackage{graphicx}
\usepackage{float}
\usepackage{graphicx, subfigure}
\usepackage{epsfig}
\usepackage{epstopdf}
\usepackage{braket}
\usepackage{hyperref}
\hypersetup{
    colorlinks=true,
    linkcolor=blue,
    filecolor=magenta,      
    urlcolor=cyan,
}
 \usepackage{url}
\usepackage{color}
\usepackage{rotating}
\usepackage{bm}

\begin{document}
\title{Topological mechanical metamaterial with nonrectilinear constraints}
\author{Natalia Lera}
\affiliation{Departamento de F\'isica de la Materia Condensada, Condensed Matter Physics Center (IFIMAC) and Instituto Nicol\'as Cabrera, Universidad Aut\'onoma de Madrid, Madrid 28049, Spain}

\author{J.V. Alvarez} 
\affiliation{Departamento de F\'isica de la Materia Condensada, Condensed Matter Physics Center (IFIMAC) and Instituto Nicol\'as Cabrera, Universidad Aut\'onoma de Madrid, Madrid 28049, Spain}
 
 \author{Kai Sun}
 \affiliation{Randall Laboratory of Physics, University of Michigan, Ann Arbor, Michigan 48109, USA}

\begin{abstract}
In this paper we study Maxwell lattices with non-rectilinear constraints, where the elastic energy is determined by the collective motion of three or more particles, in contrast to a rectilinear spring whose elastic energy only relies on the displacement of two particles attached to the two ends of the spring. Utilizing polygon-shaped constraints, we found that the Maxwell counting argument and the topological construction, based on the compatibility matrix and the equilibrium matrix, can be generalized in our models, and our elastic systems follow the same topological classification. In addition, we also found that non-rectilinear constraints offers a natural pass towards topological states with higher topological indices and multiple edge states, which can be achieved even for a simple unit cell with one degree of freedom per unit cell without enlarging the unit cell in the bulk or at the edge. 
\end{abstract}

\maketitle

\section{Introduction}
In recent studies, it has been realized that topology plays a very important role in elastic systems \cite{fiber1,KL,iop_KS,fiber4,fiber5,fiber6,fiber7,fiber8,fiber9, fiber10,fiber11,fiber12,fiber13,fiber14,fiber15,Weyl_PRL,Vishwanath,Nature_Kai, fiber18,fiber19,fiber20,ZeroDimensions_NLJV}. In particular, for systems in the verge of mechanical instability, an intriguing class of topological systems has been proposed and studied, known as Maxwell lattices, where topologically protected phonon edge modes at zero frequency (floppy modes) emerges as a result of a nontrivial topological structure of the bulk \cite{KL,iop_KS,Weyl_PRL, Vishwanath,Nature_Kai,fiber18,fiber19,fiber20}.

In this topological construction, a key step is to realize that the information about topology in these systems are not directly contained in the dynamical matrix $\mathbf{D}$, which is a standard tool for describing elastic and acoustic properties for an elastic system. To access the topological structure, it is necessary to decompose the dynamic matrix into the product of two matrices, known as the compatibility matrix $\mathbf{C}$ and the equilibrium matrix $\mathbf{Q}$, which are transpose of each other $\mathbf{C}=\mathbf{Q}^\textrm{T}$. In comparison with the dynamical matrix, these two matrices contain more information about the elastic system. On the one hand, the  $\mathbf{C}$  or  $\mathbf{Q}$ matrix uniquely determines the dynamical matrix $\mathbf{D}=\mathbf{Q} \mathbf{Q}^\textrm{T}$, up to some unimportant coefficients corresponding to the spring constants, which are irrelevant as far as topological properties in Maxwell lattices are concerned. On the other hand, however, the inverse statement is \emph{not} true. Two different elastic systems with different topological nature and different $\mathbf{C}$ or $\mathbf{Q}$ matrices can share the same dynamical matrix. In other words, as we convert the $\mathbf{C}$  or  $\mathbf{Q}$ matrix into the dynamical matrix $\mathbf{D}$, some part of the information about the elastic system is lost in this procedure. For bulk properties, i.e. the bulk phonon band structure, this information loss is irrelevant, and one can choose to work with either the dynamical matrix $\mathbf{D}$ or the $\mathbf{C}$($\mathbf{Q}$) matrix. However, for the edge phonon propagation, as well as the bulk topological structure, key information is coded in the $\mathbf{C}$  and  $\mathbf{Q}$ matrices, instead of the dynamical matrix $\mathbf{D}$. This is the fundamental reason why a more elaborate description, utilizing $\mathbf{C}$  and  $\mathbf{Q}$ matrices, becomes necessary here to characterize the bulk topology and the edge floppy modes.

In the studies about Maxwell lattices, the main efforts have been largely focused on systems with two-body central-force elastic interactions, i.e. systems composted of points connected by ideal springs. In such a system, the compatibility matrix $\mathbf{C}$ describes the relation between displacement at each site and the length variation of each spring. The equilibrium matrix $\mathbf{Q}$ dictates the relation between the tension in each spring and the total force for each site. With $\mathbf{C}$ and $\mathbf{Q}$ matrices, an effective Hamiltonian can be defined
\begin{align}
H=\begin{pmatrix}
0 & \mathbf{Q} \\ 
\mathbf{Q}^\textrm{T} & 0
\end{pmatrix}
\end{align}
The physical meaning of this effective Hamiltonian becomes more transparent when we look at the $H^2$, whose diagonal components contains the dynamical matrix $\mathbf{D}=\mathbf{Q} \mathbf{Q}^\textrm{T}$, if we ignore an overall pre-factor from the spring constant. According to the topological classification of topological insulators and superconductors, such an effective Hamiltonian belongs to the BDI class \cite{Kitaev, Classification,Classification_Proceedings}, and thus an integer topological index can be defined for a 1D system,
\begin{equation}
\nu=\frac{1}{2\pi i}\int_{-\pi}^\pi dk \textup{Tr}\left ( \mathbf{C}(k)^{-1} \frac{d \mathbf{C}(k)}{dk} \right )
\label{Eq_1DTI}
\end{equation}
where $\mathbf{C}(k)$ is the compatibility matrix defined in the $k$ space. In an electronic system, this topological index reveals the number of zero-energy excitations at the end of the 1D line. Similarly, in a mechanical system, this index dictates the number and the location of edge deformations, which costs zero elastic energy.
For the class BDI, no intrinsic topological index can be defined in 2D and 3D systems. However, one can use the idea of dimension reduction to examine topological properties of a 2D or 3D Maxwell lattices, treating a 1D line in the 2D or 3D momentum space as a 1D system. For example, in a 2D system defined in the x-y plane, each fixed $k_x$ defines a 1D line in the 2D k-space. For this 1D line, we can define the same topological index as defined above. For edges parallel to the x direction, this index reveals the number of topologically-protected zero-energy edge modes at wavevector $k_x$, as well as the location of these edge modes (i.e. top or bottom edge)~\cite{KL}. In addition, the dimension reduction concept also reveals another topological phenomena, known as mechanical Weyl modes, a topologically-protected bulk gapless mode~\cite{Weyl_PRL,Vishwanath}. In systems with elastic Weyl modes,  the topological index changes its value as we scan $k_x$ throughout the Brillouin zone. Because a topological index can only change its value through the presence of a bulk zero-energy mode, this topological structure implies that bulk gapless modes must raise, dubbed as elastic Weyl modes. These Weyl modes in general have finite wavevectors, in direct contrast to conventional bulk zero-energy modes, i.e.. acoustic phonons, whose wavevector must be zero.

In this manuscript, we explore elastic systems beyond the conventional spring-network setup, focusing on elastic forces that cannot be treated as rectilinear constraints. In Maxwell lattice, one type of non-rectilinear constraints have been considered (``tri-bond") in mechanical graphene by Socolar and coworkers \citep{SocolarKL_tribond}. In this manuscript, we focus on a different type of non-rectilinear polygon-shaped (planar or non-planar) constraints as shown in Fig.\ref{perimeter}, which has been studied in 2D foams and biological tissues  (See for example: Refs: \cite{Foams,biophysics} and references therein). Here, point particles, located at  vertices of the polygon, are joined together by the polygon. Motions of the particles will induce elastic deformation to the polygon, which costs energy similar to springs. In principle, the elastic energy for each deformation configuration can depend on the variation in perimeter, area, shape, et.~al. Here,  to demonstrate the key physical consequence of non-rectilinear constraints, we choose to focus on a specific case where the elastic energy cost only depends on the elongation/contraction of the total length of the perimeter for each polygon. Similar phenomena can also arise in other more complicated cases, which can be treated with the same approach as described below. In our model, the elastic energy is defined as
\begin{align}
E= \frac{1}{2} K \delta \mathcal{P}^2
\end{align}
where $K$ is an elastic constant, which will be set to unit for the rest part of the manuscript unless stated otherwise, and $\delta \mathcal{P}$ is the length variation of the perimeter for the polygon. To the first order approximation, we neglect any bending energy here. Later, we will study Maxwell lattices formed by rotors connected with this type of constraints,  and study topological phenomena in these systems.

In comparison with conventional Maxwell lattices, which utilizes rectilinear constraints, non-rectilinear constraints offers multi-particle interactions that directly couples the motion of three or more particles. As will be shown below, this multi-particle interactions allows more flexibility and enable us to define 
 more complicated $\mathbf{C}$ and $\mathbf{Q}$ matrices without increasing the size of a unit cell. As a result, this construction allows us to explore more generic topological states, especially those with higher topological indices and multiple zero-energy edge modes. It is worthwhile to notice here that non-rectilinear constraints are not the only pathway towards topological states with high topological indices. The same type of states can also be achieved using rectilinear springs, as long as a more complicated unit cell with multiple degrees of freedom is utilized. From the topological point of the view, the topological states obtained from this two different pathways, non-rectilinear constraints or larger unit cell, have no fundamental difference. In this sense, both approaches offers a good platform for exploring the same physics. However, as well be show below, because with non-rectilinear constrains, the high index states can be achieved with a simple unit cell with as less as one degree of freedom per cell, this approach in general can get the same physics with a smaller $\mathbf{C}$ and $\mathbf{Q}$ matrices, making analytic calculations easier. Similarly, as will be shown below, our construction also allows the realization of topological Weyl modes with extremely small/simple $\mathbf{C}$ and $\mathbf{Q}$ matrices, even down to one-by-one.

\section{perimeter}

For a polygon with $n$ vertices ($i=1,2,\ldots, n$) the displacement of the $i$th particle from its equilibrium position, $\mathbf{R}_{0i}$, can be described by the displacement vector $\mathbf{u}_i=(u_{i,x},u_{i,y},u_{i,z})$. Here, we demonstrate the most generic case, where each vertex can move in any direction in the 3D space (i.e. $\mathbf{u}_i$ is a 3D vector). However, as will be shown below, in a real elastic system (or a model system),  additional constraints can be enforced for each vertex, confining its motion to a lower dimensional space (1D or 2D). There, the same analysis is applicable. For small deformations within the linear deformation regime, the variation of the perimeter $\delta\mathcal{P}$ is a linear function of the displacement vectors, 
\begin{align}
\delta\mathcal{P}=c_{i_\alpha} u_{i_\alpha}
\end{align}
Here, repeated indices are summed over, and $\alpha=x$, $y$ or $z$. The linear coefficients $c_{i_\alpha}$ are determined by the unit vector, $\hat{n}_{ij}$, along the direction of the edges that are connected to the $i$th
\begin{align}
c_{i_\alpha}=\hat{n}_{ij,\alpha}+\hat{n}_{ij',\alpha}
\end{align}
where $j$ and $j'$ are two neighboring vertices of $i$. $\mathbf{\hat{n}}_{ij}=(\hat{n}_{ij,x},\hat{n}_{ij,y},\hat{n}_{ij,z})$ is a 3D vector component in general and


\begin{equation}
\hat{n}_{ij,\alpha}=\frac{(\mathbf{R}_{0i})_\alpha-(\mathbf{R}_{0j})_\alpha}{|(\mathbf{R}_{0i})_\alpha-(\mathbf{R}_{0j})_\alpha|}
\label{C-cartesianelements}
\end{equation}
where 
$i$ runs from 1 to $n$, $n$ being the number of sides of the polygon perimeter enumerating its vertices (sites). 
This equation is valid as long as the displacement is small. 

\begin{figure}
\epsfig{file=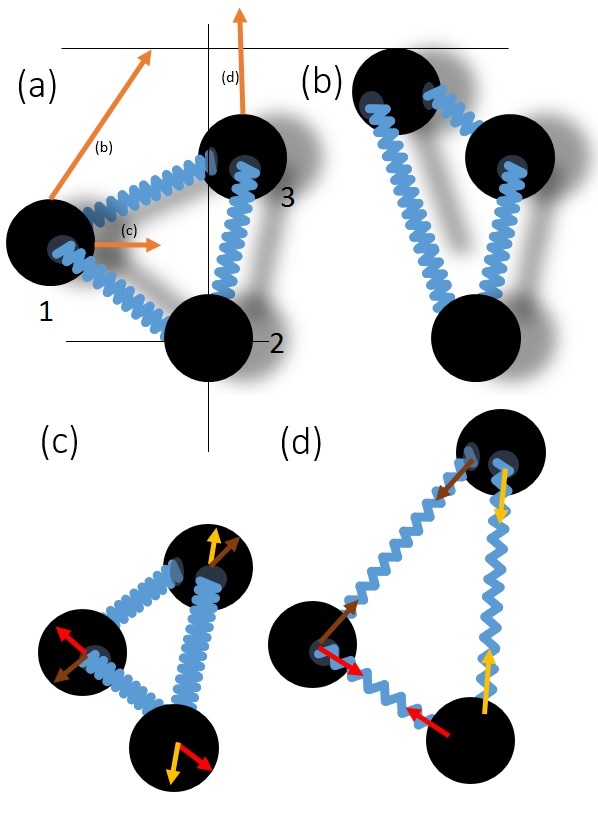,width=8cm}
\caption{(Color online) Perimeter constraint. (a) The perimeter is a closed spring connecting the points at the vertices, three in this case. (b) A different configuration of the three sites that does not modify the length of the perimeter. (c) perimeter contraction and (d) perimeter extension. 
 }
\label{perimeter}
\end{figure}

For illustration purposes, we compute the compatibility matrix of a perimeter constraint attached to three free sites in 2D, see Fig. (\ref{perimeter} (a)). The equilibrium positions in Cartesian coordinates are $\mathbf{R}_{01}=(-1,0.5)$, $\mathbf{R}_{02}=(0,0)$ and $\mathbf{R}_{03}=(0.2,1)$ for vertex 1, 2 and 3 respectively. There are a total of 6 degrees of freedom (dof), therefore, $C$ is a $1$x$6$ dimensional matrix. 
The elongation of this one perimeter is $e=\mathbf{Cu}$, where $\mathbf{u}=(\delta x_1,\delta y_1,\delta x_2,\delta y_2,\delta x_3,\delta y_3)^\textrm{T}$. The compatibility matrix can be written as $\mathbf{C}=(c_{x1},c_{y1},c_{x2},c_{y2},c_{x3},c_{y3})$, which can be easily computed by substituting in  Eq. (\ref{C-cartesianelements}), resulting $\mathbf{C}=(-1.82,0.06, 0.70,-1.43,1.12,1.37)$. Notice that displacing the site 1 along y-direction ($c_{y1}$) does stretch the perimeter slightly, compared with the contraction it will suppose to displace along x-axis ($c_{x1}$). See Fig. (\ref{perimeter} (c)).
The dynamical matrix $\mathbf{D}=\mathbf{C^\textrm{T}C}$ gives the equation of motion, $\ddot{\mathbf{u}}=\mathbf{Du}$, for this perimeter constraint. $D$ has still two zero eigenvalues, corresponding to displacements that do not modify the total perimeter like the configuration shown in Fig (\ref{perimeter} (b)).

In Fig (\ref{perimeter} (c and d)), we show the non-central forces appearing as a result of a displacement of one of the points. The tension is the same along the entire perimeter and forces appear in the direction of the polygon side.

\section{models and results}

\begin{figure}
\epsfig{file=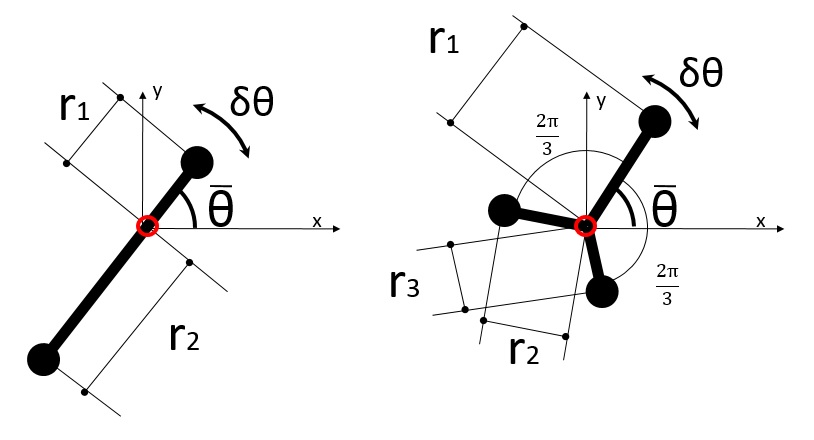,width=8cm}
\caption{(Color online) Rotors used in this paper. Both of them spin in the x-y plane and have (a) 2 and (b) 3 ends with different radius. The rotational axis is at the origin of coordinates in the figure. 
 }
\label{rotor}
\end{figure}

In this paper, all our models have the same dof per unit cell (UC). It consists on a planar rotor in the x-y plane, in which the dof is the variation of the angle referred to the angle of equilibrium. This rotor might have more than one end as shown in Fig. (\ref{rotor}). Also, the length from one site to its rotational axis may vary, while the lattice parameter, $a$, will be the unit of length.
In order to keep isostaticity, we only need one constraint per UC, leading to a scalar compatibility matrix. In this section, our perimeter constraints will be triangles.

We compute topological invariant in 1D (Eq. \ref{Eq_1DTI}). 

\subsection{1D-model}
\begin{figure}
\epsfig{file=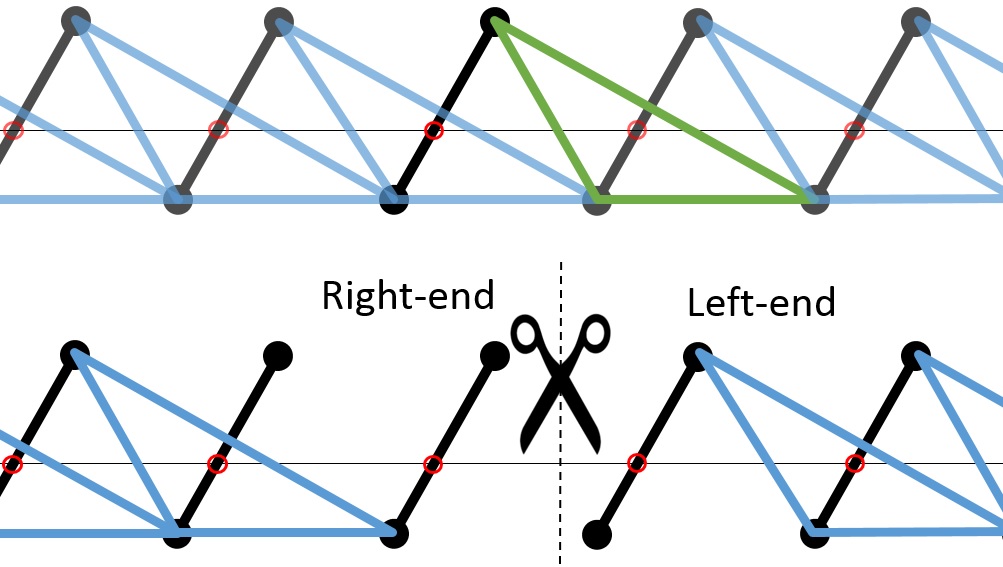,width=8cm}
\caption{(Color online) 1D Model. Top figure, the periodic system with highlighted unit cell. Bottom figure, finite system after removing two perimeters.
 }
\label{model1D}
\end{figure}

\begin{figure}
\epsfig{file=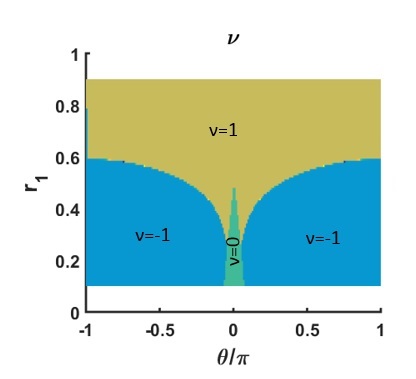,width=8cm}
\caption{(Color online) Phase diagram of 1D Model. Topological phase changes at all the gap closings, except at closing at $\bar{\theta}=\pi$.
 }
\label{model1D_PDb}
\end{figure}

\begin{figure}
\epsfig{file=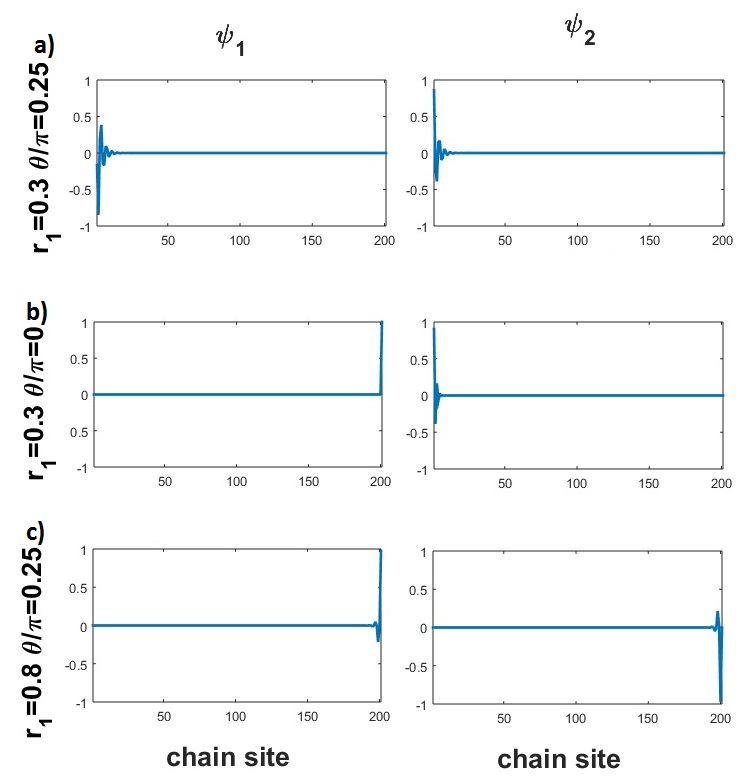,width=8cm}

\caption{(Color online) Zero modes in 1D model depicted in Fig. \ref{model1D}, with parameters representing the three different phases in Fig. \ref{model1D_PDb}, a) $\nu=-1$, b) $\nu=0$ and c) $\nu=1$. 
 }
\label{model1D_zeroModes}
\end{figure}

A simple array of rotors as the one in Fig (\ref{rotor}(a)) with two ends See Fig. (\ref{model1D}), give the following $C$-matrix structure
\begin{equation}
\mathbf{C}=c_{-1}e^{-ik_x}+c_0+c_1e^{ik_x}
\label{Cmatrix1D}
\end{equation}
where $c_{-1}$, $c_0$ and $c_1$ are real numbers as a function of the equilibrium angle $\bar{\theta}$, and rotor radius $r_1$ and $r_2$. For $\mathbf{C}$-matrix explicit form, see Appendix A.
The spectral gap is represented in Fig. (\ref{model1D_PDa}). Computing the topological invariant in 1D, for different angles of equilibrium and different values of radius with $r_2=a-r_1$, we find three different topological phases, (Fig. \ref{model1D_PDb}). Notice from Eq. (\ref{Cmatrix1D}), that three is the maximum of topological phases the model can display with a triangular perimeter and 1 dof per UC.
Finite system: cutting two perimeters in consecutive UC, we end up with a finite system (Fig \ref{model1D}). The two zero modes appear at different edges depending on the topological phase. See Fig. (\ref{model1D_zeroModes}).

\begin{figure}
\epsfig{file=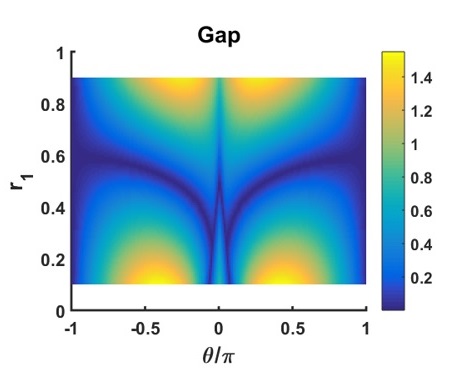,width=8cm}
\caption{(Color online) Gap structure of the 1D Model as a function of $r_1$ and $\theta$ with $a=r_1+r_2=1$ constant. 
 }
\label{model1D_PDa}
\end{figure}

\subsection{2D-model}
\begin{figure}
\epsfig{file=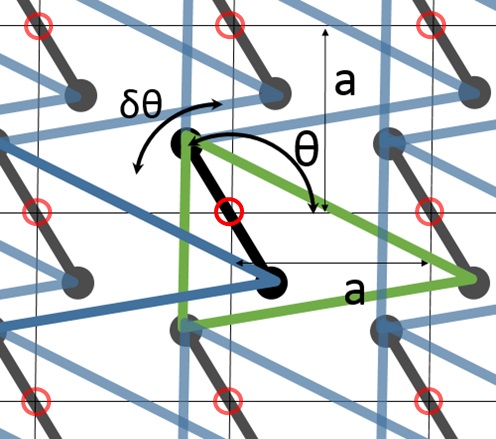,width=8cm}
\caption{(Color online) 2D model with a single rotor and one perimeter per UC, highlighted.  
}
\label{model2Da}
\end{figure}

\begin{figure}
\epsfig{file=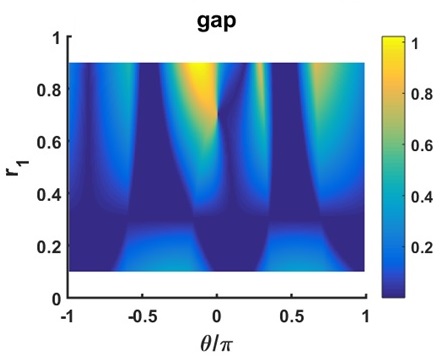,width=8cm}
\caption{(Color online) Gap value in the parameter space of this model for fixed $r_2=0.3a$ in model 2. 
}
\label{model2Db}
\end{figure}

\begin{figure}
\epsfig{file=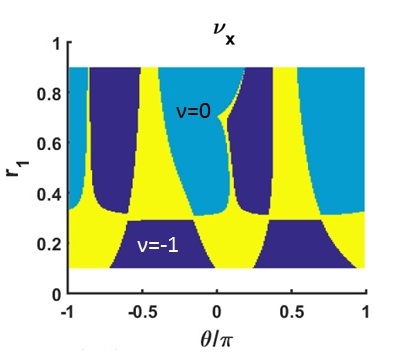,width=8cm}
\caption{(Color online) 1D topological index in x-direction in 2D model.
}
\label{model2Dc}
\end{figure}

\begin{figure}
\epsfig{file=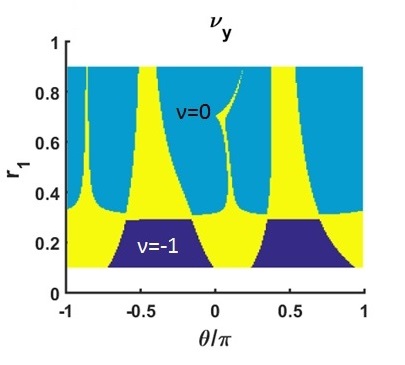,width=8cm}
\caption{(Color online) 1D topological index in y-direction in 2D model.
}
\label{model2Dd}
\end{figure}

\begin{figure}
\epsfig{file=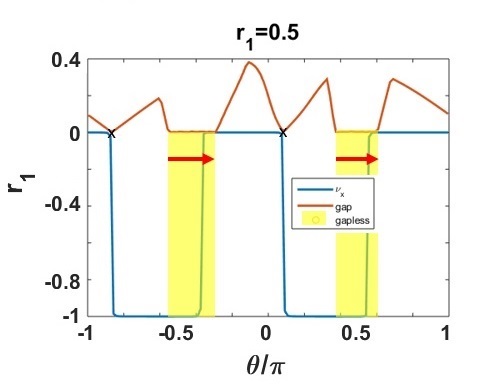,width=8cm}
\caption{(Color online) Gap and topological invariant in x-direction. The symbol $x$ signals a regular topological phase transition. Gapless areas are shadowed in yellow and crossed by arrows (see text).
 }
\label{Weyl_2Da}
\end{figure}

\begin{figure}
\epsfig{file=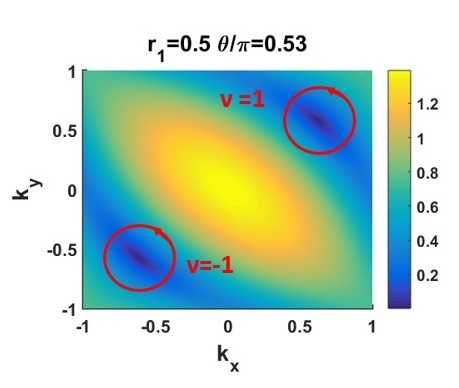,width=8cm}
\caption{(Color online) Spectrum in the 2D-BZ for a gapless system, we observe two Weyl points with opposite momenta and opposite Weyl number.
 }
\label{Weyl_2Db}
\end{figure}

This model is based in the same dof per UC promoted to two dimensions and the triangular perimeter constraint depicted in Fig (\ref{model2Da}). $C$-matrix structure is
\begin{equation}
\mathbf{C}=c_{00}+c_{01}e^{ik_y}+c_{11}e^{ik_x+ik_y}
\label{Cmatrix2D}
\end{equation}
which is explicitly written in Appendix A. In this model $r_2=0.3a$.
From $\mathbf{C}$ structure can be deduced that only 2 different phases along x will appear (two elements in $C$ with different $k_x$ phase)(Fig. \ref{model2Dc}), 2 phases along y (for the same reason)(Fig. \ref{model2Dd}) and 3 in combination (three phases in total, the maximum a triangular perimeter can display). Remember, we compute the 1D topological invariant (there is no 2D topological invariant in BDI class). In Fig. (\ref{model2Db}) is represented the total gap of the system. 
This model exhibits finite gaped areas (we will examine them below) and also gap closings leading to topological phase transitions for $\nu_x$ as can be seen in (\ref{model2Dc}).
The 1D-topological invariant is not well defined in gapless systems. For parameters inside the gapless areas, we observe Weyl points at opposite momenta in the band structure. These zero modes are topologically protected by an integer topological invariant defined on
a path that encircles them. 
While we change the system in the parameter space, Weyl points move around the Brillouin Zone (BZ). This phenomenon has been already studied in distorted squared lattice with springs \cite{Weyl_PRL,Vishwanath}.
In Fig. (\ref{Weyl_2Da}), we represent a cut of the phase diagram for fixed $r_1=0.5a$. The symbol $x$ signals a topological phase transition, with a line of zero modes, $k_x=0$. Gapless areas are shadowed in yellow and crossed by arrows. Both arrows start with the creation of a couple of Weyl points at $(\pi,\pi)$ in the $(k_x,k_y)$-BZ. The Weyl points annihilate at $(\pi,0)$ where the end of the arrows are. The spectrum in the middle of the gapless area, can be seen in Fig (\ref{Weyl_2Db}), together with the result of the Weyl topological invariant
\begin{equation}
\nu_W=\frac{1}{2\pi i} \oint_C dk \textup{Tr}\left ( \mathbf{C}(k)^{-1} \frac{d\mathbf{C}(k)}{dk} \right )
\label{Eq_1DWI}
\end{equation}

where $C$ encircles a single Weyl point.

\subsection{3D-model}
\begin{figure}
\epsfig{file=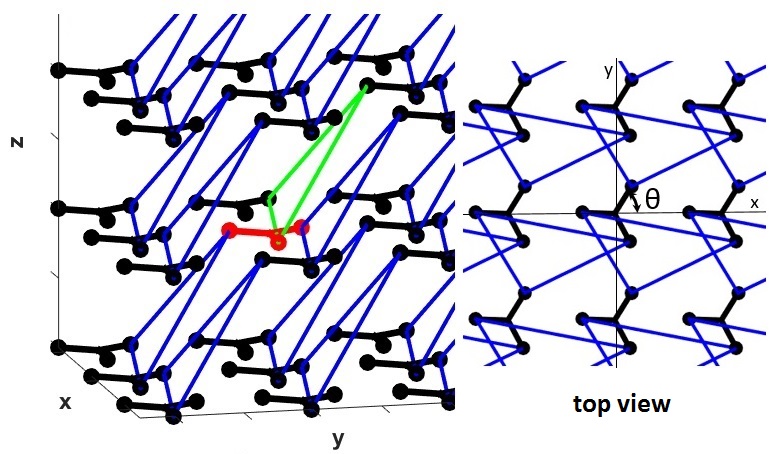,width=8cm}
\caption{(Color online) 3D Model. The periodic system with highlighted unit cell and top view of the model. 
 }
\label{model3Dab}
\end{figure}

\begin{figure}
\epsfig{file=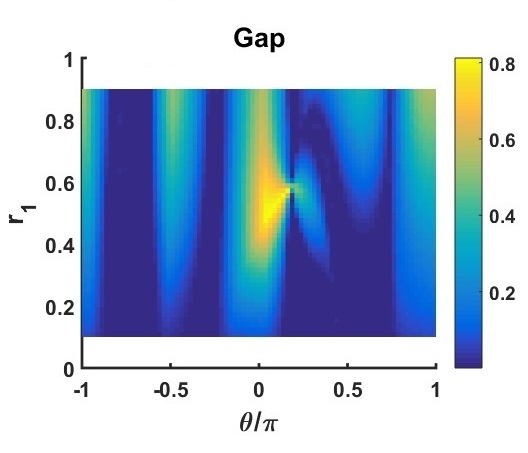,width=8cm}
\caption{(Color online) 3D Model. Gap in the parameter space.
 }
\label{model3Dc}
\end{figure}

\begin{figure}
\epsfig{file=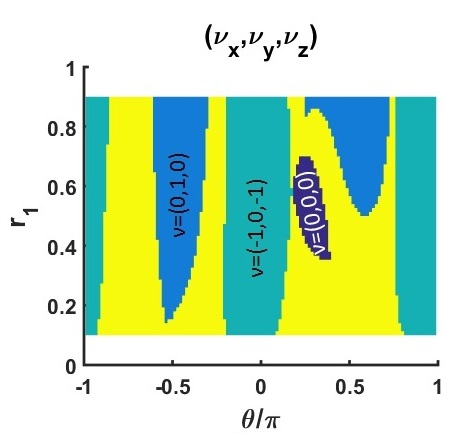,width=8cm}
\caption{(Color online) 3D Model. 1D-topological invariant in the three dimensions. In yellow, gapless areas of the parameter space, where Weyl lines appear.
 }
\label{model3Dd}
\end{figure}

This model is based in the second rotor in Fig (\ref{rotor})(b) with $r_1=r_2=r_3$, and the dof per UC is the variation of the angle in x-y plane. The triangular perimeter constraint is represented in Fig. (\ref{model3Dab}). $C$-matrix structure is
\begin{equation}
\mathbf{C}=c_{000}+c_{100}e^{ik_x}+c_{0-11}e^{-ik_y+ik_z}
\label{Cmatrix3D}
\end{equation}
where we can deduce that we can get a maximum of two phases in each direction and three in total, (Fig. \ref{model3Dd}). We observe finite areas in the parameter space of gapless systems Fig. ( \ref{model3Dc}). 

\begin{figure}
\epsfig{file=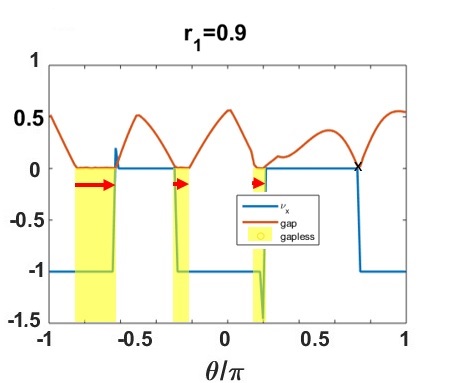,width=8cm}
\caption{(Color online) Gap and topological invariant in x-direction for $r_1=r_2=r_3=0.9$ in 3D model. The symbol $x$ signals a regular topological phase transition with a plane of zero modes. Gapless areas are shadowed in yellow and crossed by arrows (see text).
 }
\label{Weyl_3Da}
\end{figure}

\begin{figure}
\epsfig{file=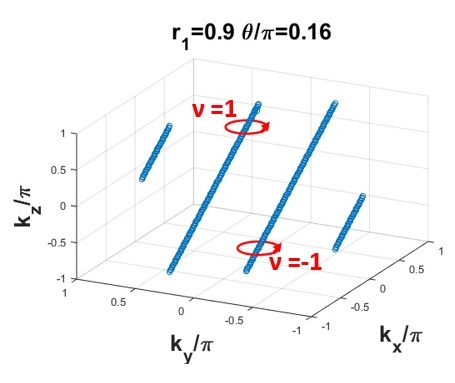,width=8cm}
\caption{(Color online) Weyl points in 3D-BZ for a gapless system, we observe two lines of Weyl points with opposite momenta and opposite Weyl number.
 }
\label{Weyl_3Db}
\end{figure}

In this areas, lines of Weyl points analogous to electronic and phononic systems arise. These Weyl lines, move around the 3D-BZ while the equilibrium parameters are changed within a gapless area. In this model, they are located at fixed $k_y$. This phenomenon has been already been studied in pyrochlore lattice with springs \cite{Weyl3D_KL}. In Fig. (\ref{Weyl_3Da}), we represent a slice of the phase diagram for $r_1=0.9a$. The symbol $x$ signals a regular topological phase transition. Gapless areas are shadowed in yellow and crossed by arrows. The three arrows start with the creation of a couple of Weyl lines at $k_y=\pi$. The Weyl lines annihilate at $k_y=0$ where the end of the arrows are. The Weyl lines inside the gapless area can be seen in Fig (\ref{Weyl_3Db}).

\section{Discussion}

In comparison with the earlier study on Maxwell lattices with rectilinear  \cite{fiber1,KL,iop_KS,Weyl_PRL,Vishwanath,Nature_Kai, ZeroDimensions_NLJV} and non-rectilinear tri-bond constraints \citep{SocolarKL_tribond}, our model reveals that polygon shape-/perimeter-based constraints can achieve the same type of diversified topological phenomena. Because the perimeter-based elastic Hamiltonian has been utilized to study various complex materials and their phase transitions \cite{Foams,biophysics}, this study may pave the road for the search for topological phenomena in bio-systems and more other complex systems. In addition, our manuscript demonstrated that the constraint that we studied here allow complicated phases to be realized within models with only one-band, and multiple edge states can arise in our model with only one degrees of freedom per unit cell and without enlarge unit cells at the edge. This observation can help understand the minimum ingredients necessary for the realization of various complex topological phases.

\subsection{quadrangular perimeters}
\begin{figure}
\epsfig{file=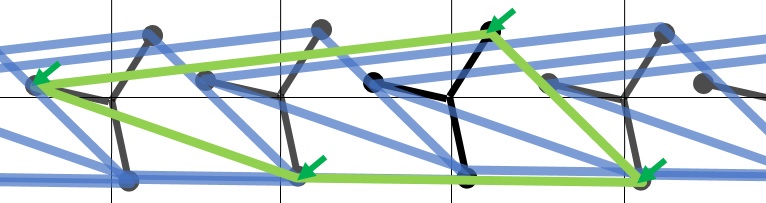,width=8cm}
\caption{(Color online) Quadrangular perimeter model in 1D with topological properties.
 }
\label{4endP_cartoon}
\end{figure}

\begin{figure}
\epsfig{file=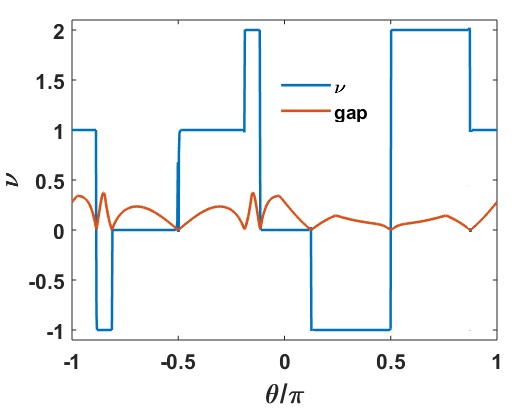,width=8cm}
\caption{(Color online) Gap and topological invariant of Quadrangular perimeter model in 1D.
 }
\label{4endP_nugap}
\end{figure}

An increasing number of phases can be obtained with perimeter polygons with larger number of edges or vertices. In the case of $n=4$, we design the model depicted in Fig. \ref{4endP_cartoon}. We use a x-y planar rotor as in Fig. (\ref{rotor} b) with $r_2=r_3=a-r_1=0.4*a$. We highlight the unit cell and signal by arrows where the 4 vertices of the perimeter are attached. Fig. \ref{4endP_nugap} shows the result of four different phases, which translates into different arrangements of zero modes for the open system. The $C$-matrix structure is,
\begin{equation}
\mathbf{C}=c_{-2}e^{-2ik_x}+c_{-1}e^{-ik_x}+c_0+c_1e^{ik_x}
\end{equation}
Notice that the four vertices are in four different unit cells, that is essential to obtain the four different topological phases. If the four vertices were in three different unit cells, only three different phases would be possible.
By removing three adjacent perimeters, we open the system and we find three zeros modes on the left for $\nu=-2$, two on the left and one on the rigth for $\nu=-1$ and one on the left and two on the right for $\nu=0$, and the three zero modes on the right for $\nu=1$. 

\subsection{bending}
If the inner angles of a planar perimeter of n sides are: $\{\alpha_1,\alpha_2,...,\alpha_n\}$. One of them can be written in terms of the other $n-1$ angles, because the sum of them all is $(n-2)\pi$. The bending energy in each vertex is proportional to the square of the complementary angle in the harmonic approximation, and the total bending energy,
\begin{equation}
E_b=\sum_i^n (\pi-\alpha_i)^2
\end{equation}
Therefore, we can define a compatibility matrix for bending $C_b$, which relates the complementary angle with the displacement. $\alpha'=C_b x$, where $\alpha'$ and $x$ are column vectors. Including both elastic and bending energies, will break the isostaticy, which is beyond the scope of this paper. 
\\

In this paper, we design a mechanical constraint which can be used to create rich topological phases and topological phenomena in Maxwell lattices even with very low number of degrees of freedom per unit cell. Such a construction greatly reduces the dimensions of the $\mathbf{C}$ and $\mathbf{Q}$ matrices. For physics observables where matrix diagonalization is required in theoretical calculations, e.g. the dispersion relation and mode shape, these smaller matrices will make analytic treatment more accessible. This non-rectilinear constrained, in polygon shape could host up to $n$ different combinations of zero modes in the open system, where $n$ is the number of sides of the polygon perimeter constraint. We showed extensively this results for triangular perimeters in 1, 2 and 3 dimensions and show the extension to larger number of sides with quadrangular perimeters in 1 dimension. Also, Weyl phases appear in wide regions of the space parameter, creating and annihilating in pairs.

$\mathit{Acknowledgments}$: NL thanks Shinsei Ryu for fruitful discussions and insights. This work has been funded by MINECO grant FIS2015-64886-C5-5-P. NL acknowledges financial support from the
Spanish Ministry of Economy and Competitiveness, through
The “Mar\'ia de Maeztu” Programme for Units of Excellence in R\&D (MDM-2014-0377), and also hospitality from the University of Michigan where part of this work was done.
KS is supported by the National Science Foundation (NSF grant EFRI-1741618) and the Alfred P. Sloan Foundation.

\bibliographystyle{ieeetr}
\bibliography{Perimeter_V1.7}

\begin{widetext}
\section{Appendix A. Compatibility matrices}
Here we show the explicit form of the compatibility matrices used along the paper. 
\subsection{Model 1D}
Compatibility matrix is just a number. First the three side lengths in equilibrium
\begin{equation}
\left\{\begin{matrix}
l_{12}=\sqrt{(r_1+r_2)^2+a^2-2a(r_1+r_2)\cos(\theta)}\\ 
l_{23}=a\\ 
l_{31}=\sqrt{(r_1+r_2)^2+4a^2-4a(r_1+r_2)\cos(\theta)}
\end{matrix}\right.
\end{equation}
The elongation in Cartesian coordinates
\begin{equation}\left\{\begin{matrix}
p_{x1}=\frac{(r_1+r_2)\cos(\theta)-a}{l_{12}}+\frac{(r_1+r_2)\cos(\theta)-2a}{l_{31}}\\ 
p_{y1}=(r_1+r_2)\sin(\theta)\left(\frac{1}{l_{12}}+\frac{1}{l_{13}} \right)\\ 
p_{x2}=\frac{a-(r_1+r_2)\cos(\theta)}{l_{12}}-\frac{a}{l_{23}}\\
p_{y2}=\frac{-(r_1+r_2)\sin(\theta)}{l_{12}}\\
p_{x3}=\frac{2a-(r_1+r_2)\cos(\theta)}{l_{12}}+\frac{a}{l_{23}}\\
p_{y3}=\frac{-(r_1+r_2)\sin(\theta)}{l_{31}}
\end{matrix}\right.
\end{equation}
and the transformation to polar variables
\begin{equation}
\left\{\begin{matrix}
c_{-1}=-r_1\cos(\theta) p_{x1}+r_1\sin(\theta)p_{y1}\\ 
c_{0}=r_2\cos(\theta) p_{x2}-r_2\sin(\theta)p_{y2}\\ 
c_{1}=r_2\cos(\theta) p_{x3}-r_2\sin(\theta)p_{y3}
\end{matrix}\right.
\end{equation}
The last three elements appear in Eq.(\ref{Cmatrix1D}), which we repeat here for convenience.
\begin{equation}
C=c_{-1}e^{-ik_x}+c_0+c_1e^{ik_x}
\end{equation}

\subsection{Model 2D}
Compatibility matrix is just a number. First the three side lengths in equilibrium
\begin{equation}
\left\{\begin{matrix}
l_{12}=a\\ 
l_{23}=\sqrt{(r_1+r_2)^2+a_x^2-2a_x(r_1+r_2)\cos(\theta)}\\ 
l_{31}=\sqrt{(r_1+r_2)^2+a_x^2+a_y^2-2a_x(r_1+r_2)\cos(\theta)-2a_y(r_1+r_2)\sin(\theta)}
\end{matrix}\right.
\end{equation}
The elongation in Cartesian coordinates
\begin{equation}\left\{\begin{matrix}
p_{x1}=\frac{(r_1+r_2)\cos(\theta)-a_X}{l_{31}}\\ 
p_{y1}=\frac{-a_y}{l_{12}}+\frac{(r_1+r_2)\sin(\theta)-a_y}{l_{31}}\\ 
p_{x2}=\frac{(r_1+r_2)\cos(\theta)-a_x}{l_{23}}\\
p_{y2}=\frac{a_y}{l_{12}}+\frac{(r_1+r_2)\sin(\theta)}{l_{23}}\\
p_{x3}=(a_x-(r_1+r_2)\cos(\theta))\left(\frac{1}{l_{31}}+\frac{1}{l_{23}}\right)\\
p_{y3}=\frac{a_y-(r_1+r_2)\sin(\theta)}{l_{31}}-\frac{(r_1+r_2)\sin(\theta)}{l_{23}}
\end{matrix}\right.
\end{equation}
and the transformation to polar variables
\begin{equation}
\left\{\begin{matrix}
c_{0}=-r_1\cos(\theta) p_{x1}+r_1\sin(\theta)p_{y1}\\
c_{01}=-r_1\cos(\theta) p_{x2}+r_1\sin(\theta)p_{y2}\\ 
c_{11}=r_2\cos(\theta) p_{x3}-r_2\sin(\theta)p_{y3}
\end{matrix}\right.
\end{equation}
The last three elements appear in Eq.(\ref{Cmatrix2D}), which we repeat here for convenience.
\begin{equation}
C=c_{00}+c_{01}e^{ik_y}+c_{11}e^{ik_x+ik_y}
\end{equation}

\subsection{Model 3D}
Compatibility matrix is just a number. First the three side lengths in equilibrium
\begin{equation}
\left\{\begin{matrix}
l_{12}=\sqrt{\left ( r(\cos(\theta)-\cos(\theta_2))-a \right )^2+\left ( r(\sin(\theta)-\sin(\theta_2)) \right )^2}\\ 
l_{23}=\sqrt{\left ( r(\cos(\theta_2)-\cos(\theta_3))+a \right )^2+\left ( r(\sin(\theta_2)-\sin(\theta_3))+a \right )^2+a^2}\\ 
l_{31}=\sqrt{\left ( r(\cos(\theta_3)-\cos(\theta)) \right )^2+\left ( r(\sin(\theta_3)-\sin(\theta))-a \right )^2+a^2}
\end{matrix}\right.
\end{equation}
where $\theta_3=\theta_2+\frac{2\pi}{3}=\theta+2\frac{2\pi}{3}$
The elongation in Cartesian coordinates
\begin{equation}\left\{\begin{matrix}
p_{x1}=\frac{r(\cos(\theta)-\cos(\theta_2))-a}{l_{12}}+\frac{r(\cos(\theta)-\cos(\theta_3))}{l_{31}}\\ 
p_{y1}=\frac{r(\sin(\theta)-\sin(\theta_2))-a}{l_{12}}+\frac{r(\sin(\theta)-\sin(\theta_3))+a}{l_{31}}\\ 
p_{x2}=\frac{r(\cos(\theta_2)-\cos(\theta))+a}{l_{12}}+\frac{r(\cos(\theta_2)-\cos(\theta_3))+a}{l_{23}}\\
p_{y2}=\frac{r(\sin(\theta_2)-\sin(\theta))}{l_{12}}+\frac{r(\sin(\theta_2)-\sin(\theta_3))+a}{l_{23}}\\
p_{x3}=\frac{r(\cos(\theta_3)-\cos(\theta))}{l_{31}}+\frac{r(\cos(\theta_3)-\cos(\theta_2))-a}{l_{23}}\\
p_{y3}=\frac{r(\sin(\theta_3)-\sin(\theta))-a}{l_{31}}+\frac{r(\sin(\theta_3)-\sin(\theta_2))-a}{l_{23}}
\end{matrix}\right.
\end{equation}
and the transformation to polar variables
\begin{equation}
\left\{\begin{matrix}
c_{000}=-r\cos(\theta) p_{x1}+r\sin(\theta)p_{y1}\\
c_{100}=-r\cos(\theta_2) p_{x2}+r\sin(\theta_2)p_{y2}\\ 
c_{0-11}=r\cos(\theta_3) p_{x3}-r\sin(\theta_3)p_{y3}
\end{matrix}\right.
\end{equation}
The last three elements appear in Eq.(\ref{Cmatrix3D}), which we repeat here for convenience.
\begin{equation}
C=c_{000}+c_{100}e^{ik_x}+c_{0-11}e^{-ik_y+ik_z}
\end{equation}

\end{widetext}

\end{document}